# MEASUREMENT AND VIBRATION STUDIES ON THE FINAL FOCUS DOUBLET AT DAΦNE AND NEW COLLIDER IMPLICATIONS*

S. Tomassini[#], INFN-LNF, Frascati, Italy

*Abstract*

A Super Flavour Factory, to be built in the Tor Vergata University campus near Frascati, Italy, will have nano-beams in order to reach a design luminosity two orders of magnitude higher than the present state of the art. The knowledge and compensation of the vibrations induced on the beams by the anthropic noise is then fundamental. The DAφNE Phi-factory at LNF, Frascati, was upgraded in the second half of 2007 in order to implement the large Piwinski angle and crab waist collision scheme [1] and in 2010 the KLOE experiment was rolled in for a new data taking and physic program [2]. A vibration measurement campaign has been performed in DAφNE to find out the actual vibration sensitivity of the final focus doublets. Vibration measurements were performed on the final focus doublet because of luminosity losses and photon beam lines instability observations. Measurement results and stabilization technique to mitigate the effects of the ground motion induced by the "cultural noise" are presented. Implications of this experience on the design and stabilization of the super flavour factory final focus doublets will be discussed.

## INTRODUCTION

On the first half of 2011, during the DAφNE commissioning and the machine tuning, in order to maximize the luminosity delivered to the KLOE experiment, vibrations of the electron-positron beams with a high peak at about 10 Hz were observed. Photon lines were affected because of the leverage amplification while the collider performances were influenced in a minor way. The vibration source was unknown but it was evident that the effect was induced on the electron-positron colliding beams mainly by the low beta quadrupoles mechanical oscillations. Quadrupole vibrations cause the colliding electron and positron beams to be offset by an amount, Δy, at the Interaction Point (IP). This leads to a luminosity loss, which can be approximated by [3]:

$$\frac{L(\Delta y)}{L_0} = \exp\left(-\frac{\Delta y^2}{8\sigma_y^2}\right)$$

In case of a vertical beam spot size σ=2.6μm, an offset Δy=2.3μm causes a luminosity drop of 10%. The vibrations of the quadrupole magnets along the beamline affect this offset at the IP by differing amounts according to the lattice optics. Vibrations of magnets far from the IP cause a small displacement, while vibrations of final focus Interaction Region (IR) quads cause an offset comparable to their movement.

___________
*Work supported by the European Commission under the FP7 Research Infrastructures project Eu-CARD, grant agreement no. 227579
#sandro.tomassini@lnf.infn.it

## MODAL ANALYSIS

The low beta magnets in DAφNE are supported by two independent long and slender aluminum cantilever structures. Each branch is supported by a Carbon Fiber Reinforced Polymer (CFRP) support inside the KLOE apparatus and outside by a metallic frame. In the central part, very close to the IP, there are no mechanical links connecting the left and right branches except for the very flexible bellows of the beryllium vacuum chamber. The calorimeter has a mass of about 500 kg. An additional lead shielding around the calorimeter is installed and has a mass of about 50 kg. QD0 mass is about 50 kg and QF1 about 15 kg. The CFRP support, that acts as a leg of the frame, has unknown characteristics as well as the lamination sequence and their mechanical properties because it was designed and built about twenty years ago and documentation does not exist anymore. Only the geometrical properties are known. An equivalent homogeneous material has been used for the Finite Element Analysis (FEA). The displacement of the head of the support was measured by means of optical instrumentation and then the Young modulus in the FEA analysis was chosen in order to get the same displacement under the same acting forces. Moreover considering the CFRP support acting as a spring with elastic constant K, it was possible to evaluate the K constant knowing the load F and the measured displacement z according the formula F=Kz. The vibration modal analysis of the Final Focus Doublet (FFD) support was performed in order to evaluate the fundamental mode shape and the corresponding natural frequency. The first vibrating mode in the vertical plane, the most critical for the collider performances, was found at about 10.4 Hz and was related mainly to the strain of the CFRP legs. The discrepancy of about 1% between the calculated and measured first mode frequency is mainly due to the approximation of the mechanical properties of the CFRP support simulated in the analysis. Results of the first four modes are listed in Figure 1.

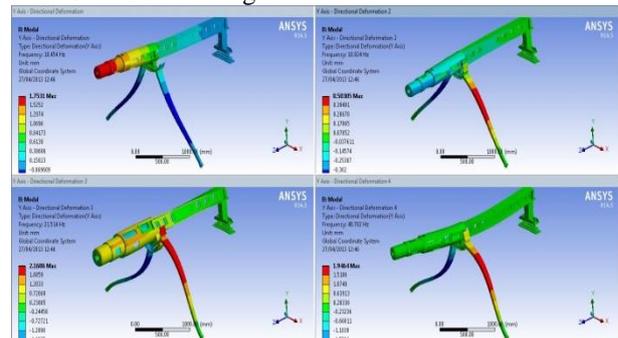

Figure 1: FFD support FEM analysis; the first four vibrating modes.

The second and the fourth mode oscillate in the horizontal plane at about 11 and 40 Hz, the third one is a torsional mode at 21 Hz, the fifth one is a forward-backward mode at 46,6 Hz and the sixth one is again a vertical mode at 54 Hz. The higher frequency modes, up to 100 Hz are only torsional.

## MECHANICAL VIBRATION MEASUREMENTS

Different campaigns of mechanical vibration measurements were performed on the FFD. Data taking setup consisted of six PCB-393B12 accelerometers and a NI PXI-4472-b board. The PXI-4472-b is 8-channels dynamic signal acquisition device, 24-bit resolution, 102.4 kS/s maximum sampling rate. PCB-393B12 are high sensitive sensors which can measure acceleration in one direction only [4]. They have a flat frequency response from 1Hz to 1 kHz. The internal noise is high at low frequencies and for this reason the signal to noise ratio is too low to enable measuring mechanical acceleration below 7 Hz. The FFT parameters used for the data analysis are following [5]:

- Window: Hanning
- Overlap: 66.67%
- Frequency resolution: 0.016Hz
- Measurement time: 20 minutes
- Averaging: Linear with 20 averages (data sets of 60 seconds averaged)

Results of the first campaign of vibration measurements are shone in Figure 5. The Root Mean Square (RMS) of the integrated displacement is reported. The measured fundamental vibrating frequency of the system was about 10.4 Hz and the integrated RMS displacement at that frequency was about 100 nm at a distance of about 1.2 m far from the interaction point (IP). The quality factor Q and the damping ratio ξ were computed according to the following formulas $Q = f_{peak}/\Delta f_{-3dB}$, $\xi = 1/2Q$. The measured damping ratio value is very low (0.001) and this fact explain why vibrating phenomena in the supporting structure tend to last for a very long period. As a consequence the structure damping ratio must be increased in order to reduce the damping time and vibration amplitude. The measured damping ratio and ground vertical Power Spectral Density (PSD) of displacement have been used as inputs for the FEM random vibration analysis in order to evaluate the real displacement of the QD0 quad that is inaccessible to be measured by means of accelerometers and is the most sensitive collider component to vibrations and then to collider performances degradation.

## RANDOM VIBRATION ANALYSIS

A random vibration analysis has been performed in order to simulate the actual response of the FFD taking into account the measured ground motion. The same FEA model used in the modal analysis has been used for the random vibration analysis and has been constrained at the same way. A global damping ratio of 0.001 has been set and the measured forcing displacement PSD with frequency ranging between 8 and 120Hz has been applied to the FFD support constraints. The results of FEM analysis were validated comparing the simulated displacement at the installation point of the accelerometer with the measured displacement. Results of the QD0 and the accelerometer installation point displacement PSD versus frequency are shown in Figure 2. The PSD displacements were integrated on the frequency span and the integrated root mean square (RMS) displacement is shone in Figure 3. At 10.4 Hz the calculated QD0 displacement amplitude is 1800 nm while at the measuring point is about 700 nm that is higher than the measured one because of the absence of damping between ground and FEM model supports.

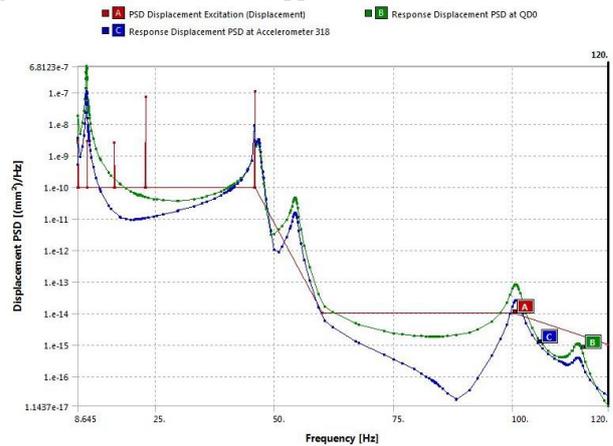

Figure 2: QD0, accelerometer installation point and forcing basement vertical displacement PSD vs frequency

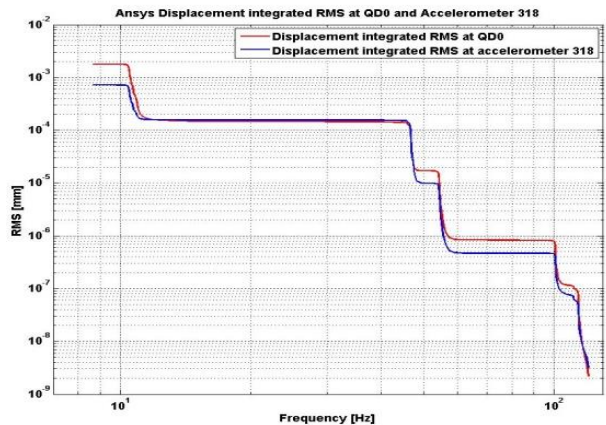

Figure 3: Integrated RMS displacement response at QD0 and accelerometer 318

## VIBRATION MITIGATION TECNIQUE

Among noted causes of the large vibrations there were water flow to the accelerator structures, heating ventilation and air conditioning, "cultural noise" in general coming from roads, plant activities and inadequate design of the FFD support structure. At the time of discovery the experimental setup as well as the collider components were installed and running. For those reasons once the problem was understood the only way to act quickly in order to mitigate the problem, was the installation of an additional passive absorber. The simplest considerable attempt consisted of some kind of "rubber-spring", acting as passively damping FFD support due to the $1/f^2$ characteristics for frequencies higher than the resonance. In order to be able to damp frequencies as low as 10.4 Hz a damper with a resonance frequency of about 7 Hz was required. This leads to a total thickness rubber of about 17mm and a rubber compression due to the magnet masses of 5.5mm. A picture of the damper is reported in Figure 4.

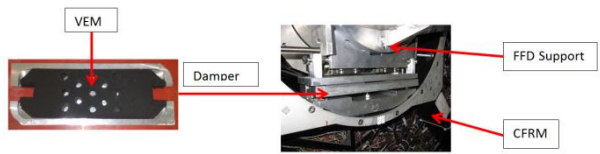

Figure 4: Passive damper and installation

The new passive vibration absorber was assembled using 3 layers of 4 mm thickness of AN-VI rubber and one layer of neoprene rubber with thickness of 5 mm. AN-VI is an especially rubber designed for vibration damping [6].

## RESULTS

A comparison between the data acquired before and after the installation of the passive damper has been done. The accelerometer used (#318) and the installation position were always the same. Mechanical vibration measurements showed a reduction of the integrated RMS displacement magnitude of about 50% at 10Hz, see Figure 5. A cross-check of results was performed comparing data from BPM measurements as shown in Figure 6. Even if data are expressed in different unit scale, an attenuation of more than 50% of the oscillation magnitude is confirmed at about 10Hz. The mechanical vibration noise seems to have small effects on the collider performances in term of geometric luminosity but nonlinear effect like beam-beam must be taken into account as well as the improvement in the photon line stability.

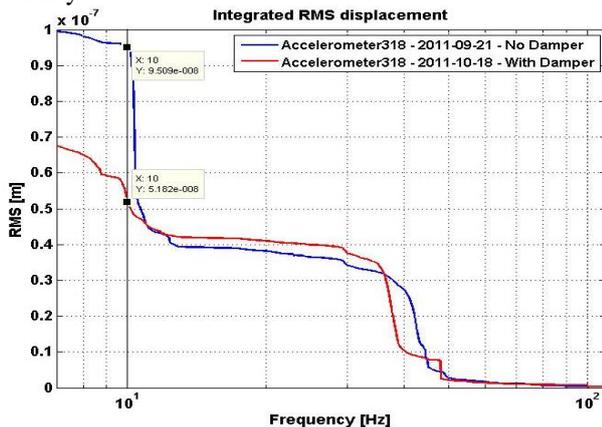

Figure 5: Measured integrated RMS displacement

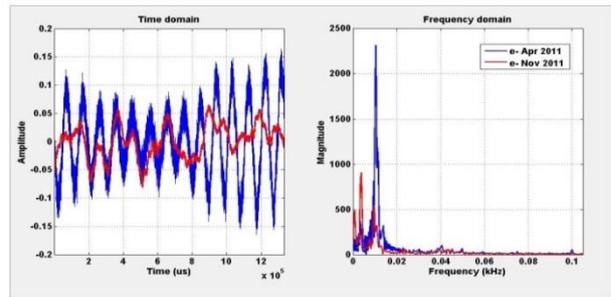

Figure 6: Electron BPM signal before and after vibration absorber installation

## CONCLUSIONS

FEM simulation and vibration measurements in the frequency range of 8-120Hz of the FFD in DAφNE have been presented. The vertical vibrations of QD0 with respect to the pit floor have two main resonant frequencies at 10.4 Hz and 46 Hz, with an Integrated RMS displacement of 1800nm and 150nm respectively. One considerable attempt to compensate FFD support resonances was a passive damping system with a resonance frequency of about 7 Hz which led to an attenuation of vibration amplitude of about 50%. Consequently the compliance (i.e. the response to a seismic horizontal wave) of such a system would be very large and the absolute alignment would not be defined with sufficient accuracy, moreover long term creep must be taken into account in the misalignment budget. To avoid excitation of inherent resonances, supports for the FFD must be designed carefully with the fundamental resonance in the vertical plane above 100 Hz. To achieve the latter statement a very stiff support with high internal damping and low mass must be designed. Composite materials are good candidates to accomplish that purpose because of their lightness, high strength and large damping. Passive vibration stabilization with a soft support is less adapted for super flavour factory accelerators, on the other hand stiff actuator supports with active position feedback using inertial reference masses must be taken into account.

## ACKNOWLEDGMENT

The Author would like to thank S. Fioravanti for the development of the data acquisition software; A. Stella for sharing BPM acquired data.